# Initialization of Neutral and Charged Exciton Spin States in a Telecom-Emitting Quantum Dot

Giora Peniakov[1,†,*], Johannes Michl[1,†], Mohamed Helal[1], Raphael Joos[2], Michael Jetter[2], Simone L. Portalupi[2], Peter Michler[2], Sven Höfling[1] and Tobias Huber-Loyola[1]

[1] Julius-Maximilians-Universität Würzburg, Physikalisches Institut, Lehrstuhl für Technische Physik, Am Hubland, 97074 Würzburg, Deutschland

[2] Institut für Halbleiteroptik und Funktionelle Grenzflächen (IHFG), Center for Integrated Quantum Science and Technology (IQST) and SCoPE, University of Stuttgart, Allmandring 3, 70569 Stuttgart, Germany

*giora.peniakov@uni-wuerzburg.de

**Abstract**

Photonic cluster states are highly entangled states that allow for photonic quantum computing and memory-less quantum repeaters. Their generation has been recently demonstrated using semiconductor quantum dots emitting at the 900 nm wavelength range. However, a similar demonstration at the communication-optimal telecom range has remained elusive. A key ingredient that is still missing is an appropriate optical excitation method. A central requirement of such a method is to allow an arbitrary spin initialization of quantum dot excitonic complexes. In this work, we report on developing such a method based on a quasi-resonant p-shell excitation for a telecom-C-band-emitting quantum dot. We show qubit writing of a neutral exciton and spin-preserving excitation of a negative trion. Using the Larmor precession of the negative trion under an externally applied magnetic field, we determine the in-plane $g$-factors of both the electron and the hole in the investigated quantum dot. In addition, we measure a lower bound on the hole coherence time, $T_2^* > 6.4$ ns, boosting its candidacy as a sound photon entangler for more advanced quantum photonic schemes.

## 1 Introduction

Quantum computing and quantum communication have captured the imagination of many who seek to exploit quantum phenomena in everyday life. For quantum computing, several technological platforms are being studied: superconductor qubits, trapped ions and atoms, photonic qubits, and others [1–4]. Photonic qubits have the advantage of low interaction with the environment, but for the use in quantum computing their wavelength has to be compatible with the processor unit [5,6]. For quantum communication, on the other hand, only photonic qubits are suitable [7]. They can be transmitted long distances while carrying quantum information encoded in their polarization [8] or in the time domain [9]. Since most of the communication is based on fused silica optical fibers, the wavelength of the photons should be chosen to minimize their losses. The telecom C-band, spanning between 1530-1560 nm, has proven to have minimal absorption in fused silica and therefore is the optimal choice.

Protocols that exploit photons for quantum communication suggest generating the photons in high-entangled graph states [10–12]. These highly entangled graph states, or cluster-states, are also the fundamental resource for photonic quantum computing. A proven way to create linear cluster-states, which can be fused to larger, more complex units [13], is by using semiconductor quantum dots (QDs) as photon sources. In 2016, Schwartz et al. showed a first demonstration of how a photonic cluster state can be generated by repetitive excitation of a QD-confined matter spin [14]. While in this demonstration the matter spin used was a dark exciton, it was later shown that a heavy hole is even a better choice, producing photons that are also indistinguishable [15]. These and other works [16–19] were based on QDs emitting at the near infrared wavelength range around 900 nm, or on neutral $^{87}$Rb atoms below 800 nm emission wavelength [20]. At the optimal range for communication, the telecom C-band, such a demonstration is still posing a challenge.

---

† These authors contributed to this work equally.

Much progress has been made in recent years developing QD photonic sources that emit at the telecom C-band range. These sources have been proven to be bright and produce single indistinguishable photons [21,22] with low multi-photon contributions. More advanced benchmarks have been achieved as well, including full spin control [23], spin-photon entanglement [24], and the generation of polarization-entangled pairs [25,26]. However, for generating a photonic cluster state, a key ingredient remained so far elusive – spin state initialization using a *non*-resonant excitation. While spin initialization using resonant excitation has already been demonstrated [23,24], this method suffers from the drawback of requiring a polarization-rejection optical setup. Since the protocol for generating cluster states utilizes polarization as the degree of freedom for qubit encoding, its realization in cross-polarization setups renders it impossible [10].

In this work, we demonstrate spin state initialization using a p-shell quasi-resonant excitation on two different excitonic complexes: the neutral exciton, $X^0$, and the negatively charged trion, $X^-$. In the first case, we initialize the $X^0$ in an arbitrary spin state using the polarization of a quasi-resonant laser pulse. We show a one-to-one correspondence between the laser polarization and the spin orientation, known in the literature as qubit writing [27,28]. In a second QD, we show polarization memory of the $X^-$ optical transition: exciting a confined electron with circularly polarized light follows an emission with the same polarization. Moreover, applying an in-plane magnetic field, we observe the Larmor precessions of both the $X^-$ and the ground state electron. These measurements allow us also to extract the in-plane electron and hole $g$-factors in the studied QDs, and to set a lower bound on the hole coherence time, $T_2^*$.

We start this paper by mathematically describing the $X^0$ and $X^-$ two-level systems, including their selection rules for optical transitions. This description is useful since much of the results in this work are based on spin precession observation. Experimentally, we detect those precessions in the photoluminescence (PL) by observing oscillations in its degree of polarization (DOP). Understanding these oscillations is therefore key in deriving the conclusions of this work.

## 1.1 The selection rules of excitonic two-level systems

In this section, we compare the spin dynamics of the $X^0$ and $X^-$, and show that they lead to a different polarization signal.

**Neutral exciton, $X^0$**

We start by defining the photon polarization basis as follows,

R/L Basis
$$\sqrt{2}|H\rangle = |R\rangle + |L\rangle; \qquad -\sqrt{2}i|V\rangle = |R\rangle - |L\rangle;$$
$$\sqrt{2}|D\rangle = \exp\left(\frac{-i\pi}{4}\right)[|R\rangle + i|L\rangle]; \quad \sqrt{2}|A\rangle = \exp\left(\frac{+i\pi}{4}\right)[|R\rangle - i|L\rangle], \qquad (1)$$

where $|H\rangle, |V\rangle, |D\rangle, |A\rangle, |R\rangle$, and $|L\rangle$ denote the horizontal, vertical, diagonal, anti-diagonal, right-hand circular, and left-hand circular polarization, respectively. The $R/L$ basis lies along the symmetry axis of the QD, $\hat{z}$, which is also the optical path direction perpendicular to the sample. In the Pauli matrices notation, the polarization $R/L$ basis coincides with the computational basis $|\pm Z\rangle$. The extra $i$ phase suggested in the $|V\rangle$ definition ensures that the relations between circular and rectilinear polarizations follow the standard polarization convention for constructing $|R\rangle$ and $|L\rangle$ from a superposition of $|H\rangle$ and $|V\rangle$. Below, we list the complete set of polarization relations in the $H/V$ basis, useful for following calculations,

H/V Basis
$$\sqrt{2}|R\rangle = |H\rangle - i|V\rangle; \quad \sqrt{2}|L\rangle = |H\rangle + i|V\rangle;$$
$$\sqrt{2}|D\rangle = |H\rangle - |V\rangle; \quad \sqrt{2}|A\rangle = |H\rangle + |V\rangle. \qquad (2)$$

In parallel, we can describe the $X^0$ two-level system using the Pauli matrices eigenstates $|+X\rangle, |-X\rangle, |+Y\rangle, |-Y\rangle, |+Z\rangle$, and $|-Z\rangle$, and set a one-to-one correspondence between them and the polarization states,

$$|H\rangle \leftrightarrow |+X\rangle; \quad |D\rangle \leftrightarrow |+Y\rangle; \quad |R\rangle \leftrightarrow |+Z\rangle;$$
$$|V\rangle \leftrightarrow |-X\rangle; \quad |A\rangle \leftrightarrow |-Y\rangle; \quad |L\rangle \leftrightarrow |-Z\rangle. \tag{3}$$

This correspondence ensures that the $X^0$ states follow the same relations defined for the polarizations in Eqs. (1)-(2). In the typical case where the QD shape deviates from a perfect cylindrical symmetry, the degeneracy of the $X^0$ two-level system is lifted, creating a fine-structure splitting (FSS), with an energy difference $E_{FSS}$. We set the direction of this asymmetry as $\hat{x}$, making the $|\pm X\rangle$ states eigenstates, and consequently redefining the reference frame for the $H$ and $V$ polarizations. Under these definitions, an arbitrary initial $X^0$ state $|\psi(t=0)\rangle = \alpha|+X\rangle + \beta|-X\rangle$ (with $\alpha$ and $\beta$ complex numbers, and $|\alpha|^2 + |\beta|^2 = 1$) evolves in time according to the Schrödinger equation:

Time evolution
$$|\psi(t)\rangle = \alpha e^{-iE_{FSS}t/2\hbar}|+X\rangle + \beta e^{iE_{FSS}t/2\hbar}|-X\rangle. \tag{4}$$

Geometrically, this evolution can be described as a rotation in the Bloch sphere around the eigenstate axis connecting the antipodal points $|+X\rangle$ and $|-X\rangle$ (see Fig. 1a).

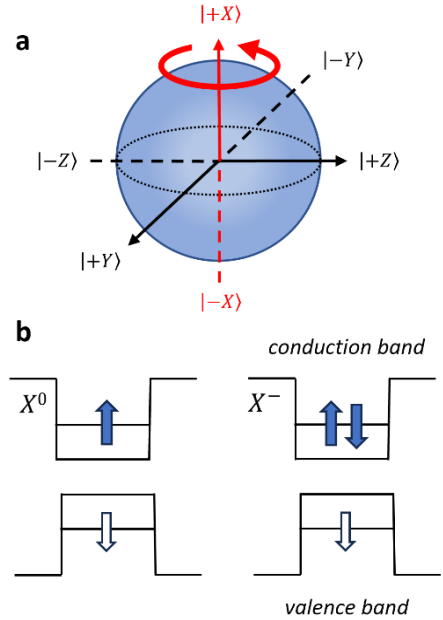

**Fig. 1**. $X^0$ **and** $X^-$. **a** Bloch sphere of the $X^0$ spin two-level system. **b** Occupation sketches of the $X^0$ and $X^-$. Without loss of generality, the hole spin direction is chosen to be down.

Since the $X^0$ is an electron-hole pair, it couples to an empty QD solely by absorbing or emitting a single photon. In general, the selection rules for optical transitions in the QD are along its symmetry axis, which we label $\hat{z}$. However, the $X^0$ single-photon coupling to the vacuum means that there is a one-to-one correspondence between the $X^0$ state and the photon polarization: a state $|\psi\rangle = \alpha|+X\rangle + \beta|-X\rangle$ couples to a photon with polarization $|P\rangle = \alpha|H\rangle + \beta|V\rangle$. Consequently, the natural precession of the $X^0$ (induced by its FSS splitting) can be observed in the polarization of the photon it emits upon optical recombination. As an example, we consider the $X^0$ prepared in the initial state $\sqrt{2}|\psi_0\rangle = |+X\rangle + |-X\rangle$ (equal to the state $\sqrt{2}|-Y\rangle$) which then evolves according to Eq. (4). Its projection probabilities on the three Bloch sphere directions as a function of time are,

$$|\langle \pm X|\psi(t)\rangle|^2 = \frac{1}{2}; \quad |\langle \pm Y|\psi(t)\rangle|^2 = \frac{1}{2}(1 \mp \cos 2\omega t)$$
$$|\langle \pm Z|\psi(t)\rangle|^2 = \frac{1}{2}(1 \mp \sin 2\omega t) \tag{5}$$

where we define the frequency as $\omega = E_{FSS}/2\hbar$. These projection probabilities explicitly represent the polarization of the emitted light in the experiment,

$$\langle H/V\rangle = |\langle \pm X|\psi(t)\rangle|^2 \quad \langle D/A\rangle = |\langle \pm Y|\psi(t)\rangle|^2 \quad \langle R/L\rangle = |\langle \pm Z|\psi(t)\rangle|^2. \tag{6}$$

Then, the measured degree of polarization (DOP) in each basis is defined by,

$$DOP_X = \frac{\langle H\rangle - \langle V\rangle}{\langle H\rangle + \langle V\rangle}; \quad DOP_Y = \frac{\langle D\rangle - \langle A\rangle}{\langle D\rangle + \langle A\rangle}; \quad DOP_Z = \frac{\langle R\rangle - \langle L\rangle}{\langle R\rangle + \langle L\rangle}. \tag{7}$$

Plugging the projection probabilities of Eq. (5) into the DOPs, yields,

$$DOP_X = 0; \quad DOP_Y = -\cos\frac{E_{FSS}\,t}{\hbar}; \quad DOP_Z = -\sin\frac{E_{FSS}\,t}{\hbar}. \tag{8}$$

**Charged exciton (trion), $X^{\pm}$**

The case of the trion, $X^{\pm}$, is slightly different. It comprises three charge carriers, where an electron-hole pair recombination renders the QD with a single charge. The spin of this charge is often called the ground-level spin, while the spin of the trion itself, governed by the unpaired charge carrier, is called the excited-level spin [29]. For example, in the case of the negative trion, $X^-$, the ground-level spin is an electron while the excited-level spin is the $X^-$ itself. However, since both of the $X^-$ electrons reside in their lowest s-shell energy level, they form a singlet and therefore do not contribute to the overall angular momentum. The unpaired hole, then, governs the spin of the entire excited level.

First, we assume the $X^{\pm}$ is prepared in the state $\sqrt{2}|\psi_0^T\rangle = |T_+\rangle + |T_-\rangle$, similarly to our previous discussion on the $X^0$. In this notation, $|T_{\pm}\rangle$ are the equivalent states to the neutral exciton's $|\pm X\rangle$. They follow the states relations specified in Eq. (1)-(2). For example, $\sqrt{2}|T_+\rangle = |T_\uparrow\rangle + |T_\downarrow\rangle$, where the states $|T_{\uparrow/\downarrow}\rangle$ correspond to the computation basis and equivalent to the $X^0$'s $|\pm Z\rangle$. We further assume that applying an external magnetic field splits the $|T_{\pm}\rangle$ states by an amount $\delta$, playing a similar role as the excitonic FSS. As a function of time, the state $|\psi_0^T\rangle$ evolves into

$$\sqrt{2}|\psi^T(t)\rangle = e^{-i\omega t}|T_+\rangle + e^{i\omega t}|T_-\rangle, \tag{9}$$

where $\omega = \delta/2\hbar$. As before, the selection rules for optical transition are along the $\hat{z}$ direction and read $|T_\uparrow\rangle \mapsto |\uparrow\rangle|R\rangle$ and $|T_\downarrow\rangle \mapsto |\downarrow\rangle|L\rangle$, where $|\uparrow/\downarrow\rangle$ are the single charge-carrier spin projections parallel to $|T_{\uparrow/\downarrow}\rangle$. Upon electron-hole recombination, the spin of the rendered charge carrier and the emitted photon polarization form the joint superposition,

$$|\psi(t)\rangle = \frac{1}{2}[(e^{-i\omega t} + ie^{+i\omega t})|\uparrow\rangle|R\rangle + (e^{-i\omega t} - ie^{+i\omega t})|\downarrow\rangle|L\rangle]. \tag{10}$$

Calculating the projections $|\langle H,V,D,A,R,L|\psi(t)\rangle|^2$ and then the resulting DOPs, we obtain,

$$DOP_X = 0; \quad DOP_Y = 0; \quad DOP_Z = -\sin\frac{\delta t}{\hbar}. \tag{11}$$

Note that assuming a different selection rule, say, $|T_{\pm Y}\rangle \mapsto |D/A\rangle|\pm Y\rangle$, would produce a different result from Eq. (11).

Comparing Eq. (8) and Eq. (11), one sees a substantial difference between the $X^-$ the $X^0$ cases: while for the $X^0$, oscillations in the emitted polarization are expected both in the $Z$ and $Y$ bases, for the $X^{\pm}$, they are expected only in $Z$. The main difference between the two cases stems from the difference in the final states of these two optical transitions: while the $X^0$ decays into vacuum, the $X^-$ decays into another two-level system - the electron.

In the next section, we will examine this difference experimentally. We will show that we can write an arbitrary superposition onto the $X^0$ state and prove it, while on the $X^-$ we can only initialize in two spin states, $|T_{\uparrow/\downarrow}\rangle$.

## 2 Results

### 2.1 Neutral exciton, $X^0$

*Time-independent spectroscopy*

Fig. **2**a shows a polarization-resolved micro-photoluminescence (μ-PL) spectrum of a selected QD1. Several spectral lines of discrete optical transitions portray a typical QD emission at 1530-1540 nm. For excitation, we use a 1064 nm above-bandgap continuous-wave (CW) laser. We identify the high-energy line at ~1535 nm as the neutral exciton, $X^0$. This identification is based on (i) observation of a polarized FSS of $36.8 \pm 2.6$ μeV, (ii) its high-energy spectral position, compatible with the characteristic low binding energy of an $X^0$ relative to other excitonic complexes, and (iii) occupation statistics under a power series measurement: an increase in the excitation power reveals the $X^0$ line as among the first to appear, indicating on its low-number-of-particle structure. Based on the FSS measured polarization, we redefine the $H$ and $V$ polarization directions of our setup to match the orientation of the QD.

To identify the quasi-resonances of the QD, we perform a photoluminescence excitation (PLE) scan. For that, we use a CW excitation laser with a tunable wavelength between ~1460-1527 nm. In energy terms, this range covers about 37 meV: from 4.2 to 41.5 meV above the $X^0$ emission. Fig. 2b shows a part of this scan where we detect a quasi-resonance at approximately 10.3 meV. Similarly to the μ-PL in Fig. 2a, we measure this PLE in two polarization conditions: one where both the excitation and the detection are set to $H$, and one where both are set to $V$. We make several observations:

1. The $X^0$ emission energy in the PLE is shifted compared to its emission under above-bandgap excitation (where $\Delta E_{PL} = 0$). This shift, relative to the central wavelength of the quasi-resonance, is approximately $\Delta E_{PL} = -0.8$ meV. A possible explanation for that is the Stark effect caused by charge accumulation in the vicinity of the QD: we conjecture that changing the excitation wavelength from above-bandgap to below-bandgap (which the PLE scanning laser is) changes the charge environment of the QD. As a result, different electric fields are induced in the QD, modifying confined energy levels, and spectrally shifting PL. Excitation power, on the other hand, only weakly affects the spectral position of the lines (not shown). This observation is reminiscent of the photoelectric effect and strengthens our conjecture about wavelength-dependent varying charge environment.
2. The quasi-resonance has a diagonal-shape profile with width and height (along $\Delta E_{PL}$ and $\Delta E_{exc}$) of roughly 0.5 and 0.6 meV, respectively. Its slope is, therefore, $1.22 \pm 0.02$. Note that the spectral width of 0.5 meV is an order of magnitude larger than the resolved linewidth of the spectral line, which we measure around 80 μeV. As of the time of writing, the origin of the quasi-resonance diagonal shape remains unclear for us.
3. The quasi-resonance also reveals an FSS of the p-shell. As a function of the laser energy only, $\Delta E_{exc}$, this splitting is $49 \pm 4$ μeV (vertical direction in Fig. 2c). Perpendicular to the diagonal, it is $31 \pm 4$ μeV (fine-dash line). Examples of similar values for p-shell FSS can be found in Ref. [30].

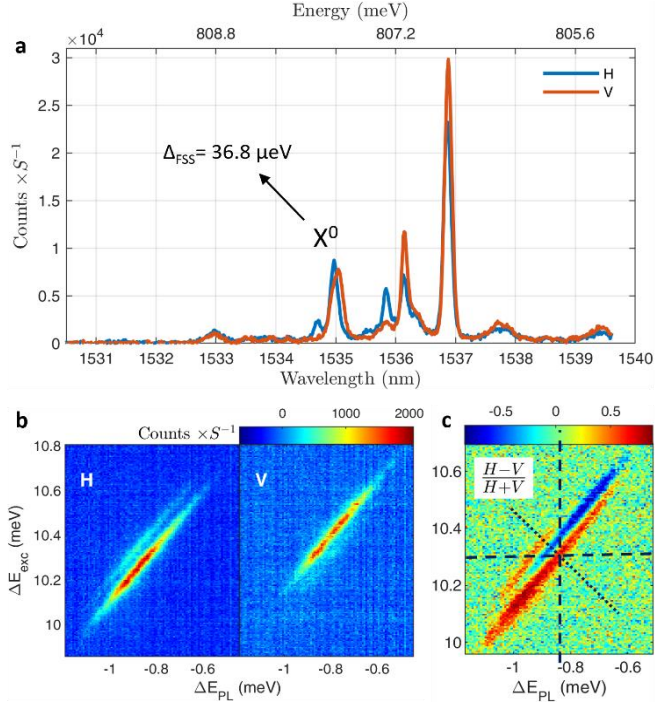

**Fig. 2. $X^0$ PL and PLE. a** PL spectrum of QD1 projected on $H$ and $V$ polarizations. **b** PLE of the $X^0$, concentrated on its p-shell quasi-resonance found around 10 meV above the $X^0$ emission. The quasi-resonance is shown under two polarization conditions: in the left pane, both excitation and detection are $H$; in the right pane, both are $V$. **c** DOP of Fig. 2b. Values close to 1 (-1) are marked red (blue) corresponding to $H$ ($V$) polarization. Dashed lines are guides to the eye showing different directions of calculated energy splitting (see text).

*Polarization-resolved time-dependent spectroscopy*

To demonstrate qubit writing on the $X^0$, we measure its time-dependent emission after pulsed laser excitation. The laser pulse duration is around 15 ps, and its wavelength 1514.4 nm, matching the $X^0$ quasi-resonance presented earlier in Fig. 2b. The emission is recorded using superconducting nanowire single-photon detectors (SNSPDs) with a response time jitter of around 28 ps. To control the polarization of the excitation and detection channels, we use liquid crystal variable retarders (LCVRs). By periodically changing the polarization of each channel between $H, V, D, A, R,$ and $L$, we recorded $6 \times 6 = 36$ time-traces of different excitation-detection combinations. In Fig. 3, we present only six of them, best illustrating the $X^0$ selection rules for optical recombination (the full set of 36 appears in the supplementary information). Since here we are mostly interested in the oscillation dynamics, we normalized each time-trace by the time-integrated signal (i.e. by the area beneath each decay). We often refer to this measurement as a "polarization-resolved lifetime measurement". The lifetime here was measured $\tau = 1015 \pm 10$ ps.

In panels (a-c), we excite the $X^0$ using $R$-,$D$-, and $H$-polarized pulses and project the signal on the corresponding bases, $R/L$, $D/A$, and $H/V$. In the $R/L$ and $D/A$ bases, clear oscillations of the $X^0$ are observed. A phase shift of $\pi$ is evident between the oscillations in the co- and cross-polarizations (blue and red curves). The lower panels of (a-c) show the DOP of the oscillations, from which we extract the oscillation period, $T = 114 \pm 1.5$ ps. This value corresponds to an energy splitting of $\Delta E = 2\pi\hbar/T = 36.17 \pm 0.47$ μeV, in a good agreement with the spectrally measured FSS mention earlier, $36.8 \pm 2.6$ μeV.

In the $H/V$ basis, we observe only a mild residual undulation of the polarization. From theoretical considerations, one would expect to observe no oscillations in this basis at all. First, because exciting with $H$ or $V$, we assume to initialize the $X^0$ in one of its eigenstates, which stays constant in time (Eq. (4)); second, even for an evolving state, its projection on the eigenstate basis ($H/V$) should stay constant (also Eq. (4)). (Geometrically, a rotation around any axis in the Bloch sphere induces a constant projection on it).

We suggest the following explanations for the residual undulations: For the latter point, we assume that the polarization projection alignment of our setup does not perfectly agree with the $X^0$ orientation. For the former point, we speculate that we fail to initialize the $X^0$ exclusively in one of its eigenstates because we inevitably excite a small portion of the orthogonal state. This is caused by the diagonal shape of the quasi-resonance discussed earlier (Fig. 2c): it forces any given excitation wavelength to couple both to the $H$ and $V$ resonance components, blocking the possibility of isolating only one of them. This renders the polarization of excitation as the only remaining degree of freedom to exclusively select one of the excited-state eigenstates. A contrary scenario can be found in Ref. [27], performed on QDs emitting at the 900 nm range. There, the quasi-resonance profiles are flat, and hence both polarization and wavelength are useful knobs for selective excitation.

We make an additional remark regarding the *visibility* of the oscillations in Fig. 3. The visibility is defined as the maximal value of the DOP during the oscillations. In the $R/L$ basis, for example, it is bounded by 0.2, far from the ideal case of 1. We attribute this reduction, at least partially, to the response time uncertainty of our measurement setup (time jitter), which is around 28 ps. Since this value is on the same time scale as the $X^0$ oscillations, the convolution of the system's response with the measured signal reduces its visibility. (see detailed analysis in Ref. [31]).

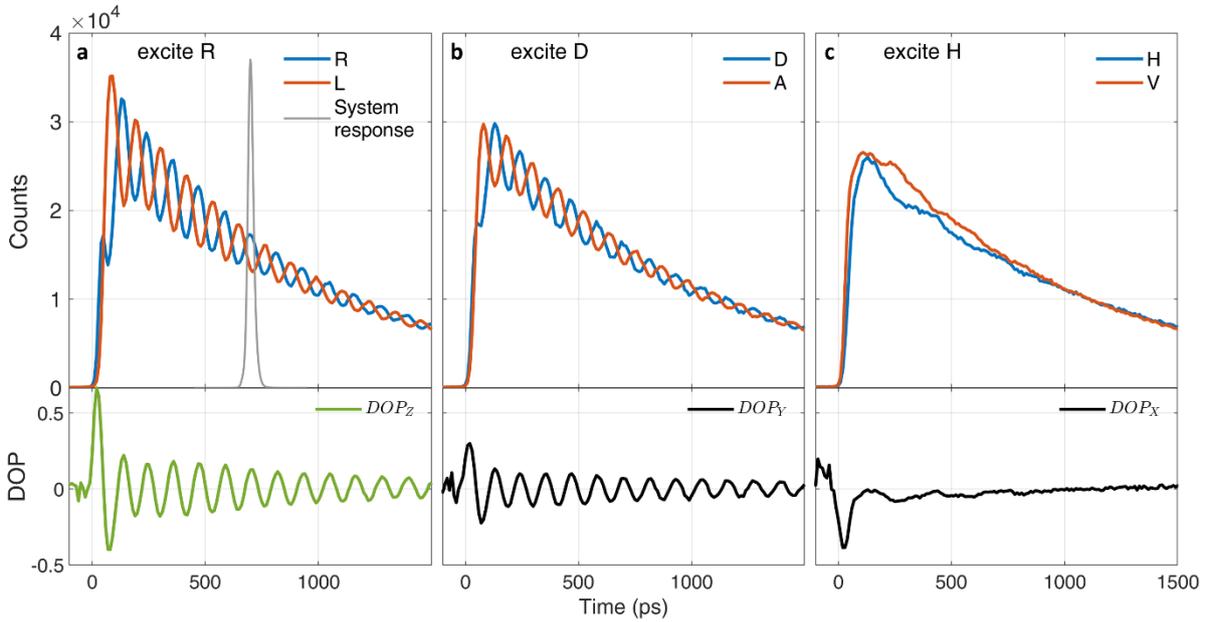

**Fig. 3. Temporal evolution of the $X^0$ from the moment of its creation. a-c** (top) Polarization-resolved evolution when the $X^0$ is excited with $R$-, $D$-, and $H$-polarized pulses and measured in the $R/L$, $D/A$, and $H/V$ bases, respectively. Each time-trace is normalized by the area beneath it. The system response is presented as a reference, where its amplitude is scaled to match the signal. **a-c** (bottom) The degree of polarization (DOP) per basis, defined as $(P - \bar{P})/(P + \bar{P})$, where $P$ and $\bar{P}$ are orthogonal polarizations. The circular DOP of (a) is plotted in green to match another use of the data later in Fig. 4.

To further analyze the one-to-one correspondence between the excitation polarization and the generated $X^0$ spin state, also known as qubit writing, we use the same data set of 36 polarizations but focus only on those where the detection is in the circular basis, $R/L$. In Fig. 4, we show the circular DOP of the $X^0$ decay upon excitation with the six polarizations, $R, L, D, A, H$, and $V$, corresponding to the $DOP_Z$ calculated in Eq. (7). To qualitatively analyze the qubit writing, we compare the phases of the $DOP_Z$ signal with different excitation polarizations. We expect a $DOP_Z$ signal with a $\pi/2$ phase shift for initializing in $|\pm Z\rangle$ and a $\pi/4$ phase shift compared to this when initializing in $|\pm Y\rangle$. Initialization in $|\pm X\rangle$ should not show any oscillations, see Eq. (8). Exciting with $R$ ($L$) should initialize $|+Z\rangle$ ($|-Z\rangle$) while excitation with $D$ ($A$) should initialize $|+Y\rangle$ ($|-Y\rangle$). Four vertical dashed lines between panels (a) and (b) mark four adjacent peaks of the four oscillations when excited with $R, L, D, A$. The color of each line matches the corresponding polarization. One can see that the phase spacing between the four polarization is roughly $\pi/4$, as expected.

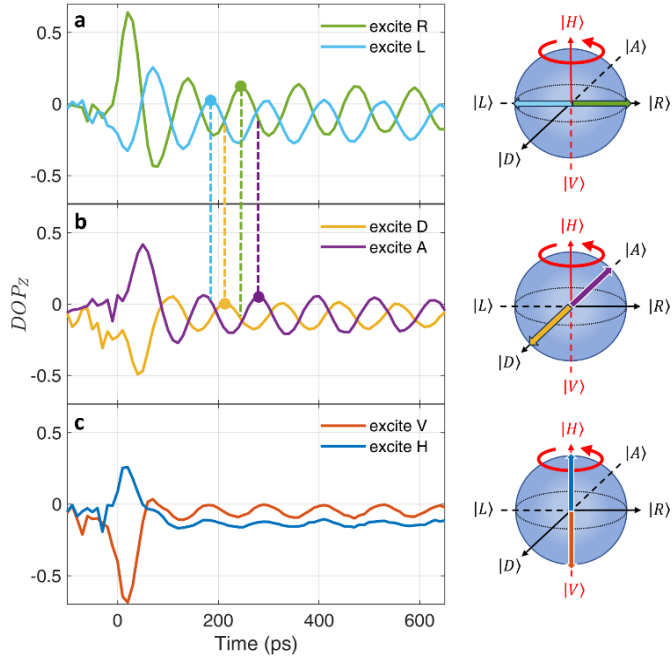

**Fig. 4. Qubit writing. a-c** Circular DOPs ($DOP_Z$) of the $X^0$ emission which follows an excitation with $R/L$, $D/A$, and $H/V$ polarized pulses. Next to every panel, a color arrow marks the $X^0$ ideal state initialization in a Bloch sphere. The dashed colored vertical lines connecting panels **a** and **b** mark the four phases as determined by the maxima point of adjacent oscillation peaks.

A quantitative measure of the four phases of $R$, $D$, $L$, and $A$ reveals a deviation from a perfect $\pi/4$ spacing (see Supplementary Fig. 4). In addition, the $H$ and $V$ time traces reveal residual oscillation where a flat dependence is expected, similar to the imperfection in Fig. 3c. A probable reason for these imperfections is the challenge of ensuring that the six laser polarizations maintain orthogonality by the time they are absorbed in the QD. While prepared orthogonal in the lab frame, they can undergo skewing after passing several optical elements before reaching the QD.

Despite these imperfection, our results provide a proof-of-concept that the $X^0$ spin state can be arbitrarily prepared anywhere on the Bloch sphere. For a practical application, one can fine-tune each of the excitation polarizations to perfectly match the targeted states. While compromising on polarization orthogonality in the lab frame, such fine-tuning would still fullfil its task. As can be seen in the Supplementary Fig. 3, the polarization can be fine-tuned to minimize oscillations in $V$ in the expense of $H$, opposite to the case presented in Fig. 4c. Thus, when combining those two excitation optimizations one would be able to write either H or V.

## 2.2 $X^-$ Larmor precession in a lifetime measurement

In the following section, we concentrate on a different spectral line – the negative trion, $X^-$, which we found in a different QD, QD2. Fig. 5a presents the spectrum of QD2 and PLE of the $X^-$. In contrast to QD1, here, we show the PL directly under quasi-resonant and not under above-bandgap excitation. Under these conditions, the $X^-$ appears in the spectrum as a single line, emitting around 1541.6 nm. Similarly to the $X^0$, the PLE scan of the $X^-$ reveals a diagonal shape of its quasi-resonance, but with a slope of $0.82 \pm 0.06$. Our identification of the spectral line as an $X^-$ is based mostly on its time-traced behavior presented in the next section. As a first hint, however, we observe no evidence of FSS. Observing FSS would immediately rule out a trion state (see Kramers' theorem, e.g. in Ref. [32]). Secondly, since a bare quasi-resonant laser pumps electron-hole pairs into specific energy levels, we assume that it can maintain a steady state population of only an exciton or a trion[30]. Power series performed with the quasi-resonant laser bring us to the same conclusion (see Supplementary Fig. 1b). We will return to the $X^-$ identification by the end of the text where we thoroughly discuss how we determined its sign.

For the $X^-$, direct qubit writing is not possible, as explained in the introduction. However, it may have polarization memory that is useful in initializing a matter spin, at least along the selection rules direction, $\hat{z}$ [32]. We say that a given optical transition has polarization memory if exciting it with $R$-polarized light results in predominantly $R$-polarized emission (with the $L$-component suppressed), and vice versa. Indeed, in Fig. 5b we show how exciting the $X^-$ in QD2 using its quasi-resonance produces this effect. It is quantified in a similar way as the circular DOP, $DOP_Z = (R - L)/(R + L)$, but time-integrated over the entire signal. Using this definition, we find the polarization memory of the $X^-$ to be $0.66 \pm 0.07$, where we averaged over excitation with $R$ and $L$ polarizations.

In Fig. 5c, we show similar measurements, but where we add an external in-plane magnetic field. The oscillations observed in the DOP for different magnetic fields can be directly linked to the Larmor precession of the excited-level spin. For the $X^-$, it is determined by the unpaired hole. Fitting the oscillation frequency versus the magnetic field using a linear curve, we extract the in-plane hole $g$-factor, $|g_h| = 0.593 \pm 0.002$ (Fig. 5d). This value qualitatively agrees with previous works of Belykh et al. where they reported $|g_h| = 0.64 \pm 0.08$ [33,34].

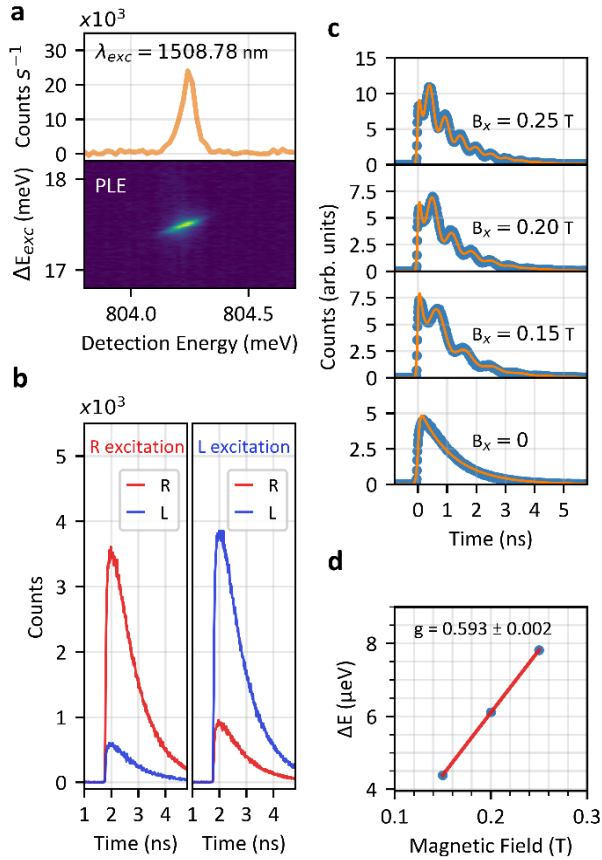

**Fig. 5. $X^-$ under quasi-resonant excitation. a** Spectral signature of the $X^-$ and its PLE excitation. **b** Polarization memory. The $X^-$ transition is excited once in $R$ and once in $L$ while detected in the co- and cross-polarizations. The integrated $DOP_Z$, also referred to as polarization memory, is measured $0.66 \pm 0.07$. **c** Time-trace of the $X^-$ decay at different values of an in-plane magnetic field. The oscillation frequency increases with higher fields and is extracted by fitting the time traces. **d** Linear fit of the energy splitting resulting from **c** vs. magnetic field. From the linear slope, a $g$-factor absolute-value of $0.593 \pm 0.002$ is calculated.

To extract the spin coherence time of the excited state, i.e. the unpaired hole in the $X^-$, we resolve the oscillation polarization in its lifetime decay. A measurement under $B_x = 200$ mT is summarized in Fig. 6. This data set constitutes an analogous set to the one presented earlier for the $X^0$ in Fig. 3: it shows six time traces, divided into three polarization bases, $R/L$, $D/A$, and $H/V$. Each panel (a-c) shows excitation in one polarization, and signal detection in the corresponding co- and cross-polarizations. The full data set of 36 combinations of excitation and detection polarizations are presented in the Supplementary Fig. 5. Note that the only conceptional difference between the $X^0$ and $X^-$ cases is the mechanism lifting the spin degeneracy: while in the $X^0$ case, it is the FSS, here, it is the in-plane magnetic field.

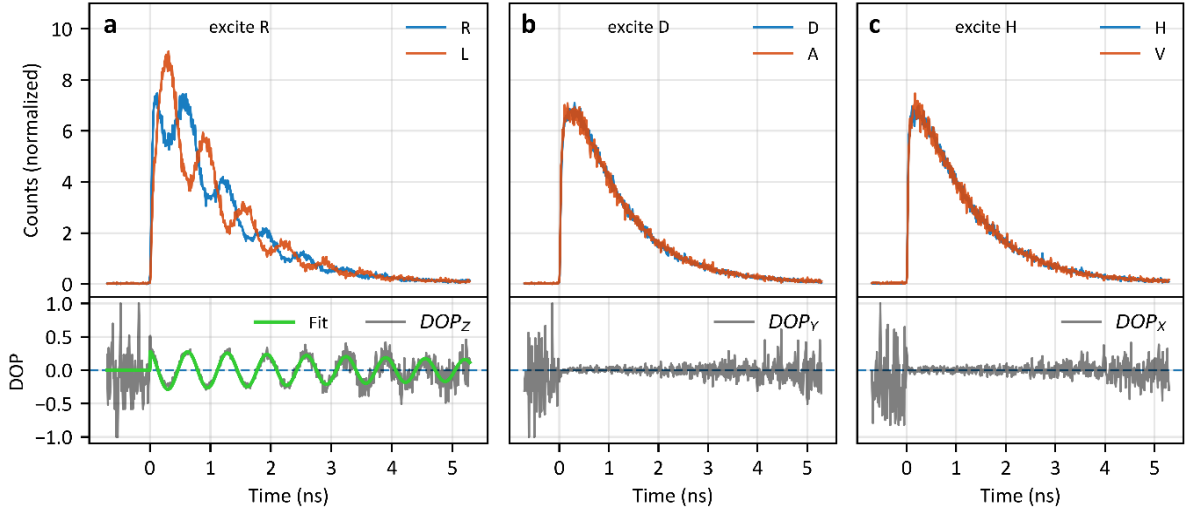

**Fig. 6. Time-trace of the $X^-$ decay after pulsed excitation** (same figure structure as Fig. 3 for the $X^0$). **a-c** (top) Excitation with $R$, $D$, and $H$ polarizations while detection in the respective co- and cross-polarizations. **a-c** (bottom) DOP. As opposed to the $X^0$, oscillations are observed only in the $R/L$ basis, not in $D/A$. Fitting the $DOP_Z$ yields a hole coherence time of $8.8 \pm 2.3$ ns.

Comparing Fig. 6 with the $X^0$ results in Fig. 3, we see a clear difference in the diagonal basis, $D/A$. While in the $X^0$ case, spin precession is clearly observed in this basis, for the $X^-$, it is not. This difference is a direct consequence of the $X^-$ having a different final state after optical recombination than the $X^0$: while the $X^0$ decays into vacuum, the $X^-$ decays into a two-level spin system formed by a single electron. Theoretically, this difference is evident comparing Eq. (8) and (11).

Finally, we can use the $X^-$ oscillation decay to extract the hole coherence time, $T_2^*$. We do that by fitting the DOP of the oscillations, presented in the lower panel of Fig. 6a, with the function $e^{-t/T_2^*}\cos \omega t$. We find $T_2^* = 8.6 \pm 2.2$ ns. The relatively large error stems from the increasing statistical noise of the signal as it drops beyond the optical transition lifetime, $\tau = 1.135 \pm 0.022$ ns. Therefore, we suggest the lower margin of this value as a lower bound to the real coherence time, $T_2^* > 6.4$ ns. To the best of our knowledge, this bound is the highest reported so far for holes in telecom-emitting QDs.

## 2.3 $X^-$ Larmor precession in a $g^{(2)}(\tau)$ correlation measurement

So far, we showed how a trion decay can reveal its *excited-level* spin precession. For the $X^-$, that was the hole. Here, we present a two-photon correlation measurement ($g^{(2)}(\tau)$) that allowed us to simultaneously observe both the excited- *and the ground-level* spin precession. For that, we use a CW laser to quasi-resonantly excite the optical transition, same as in the previous section. The laser polarization, as well as the two-photon projection polarizations, were all set to $R$. A similar measurement for InAs QDs emitting at the 900 nm range was recently reported in Ref. [35].

An example for this measurement, taken with an in-plane field of $B_x = 40$ mT, appears in Fig. 7a. Note that the data displayed here is already normalized to a fit of the antibunching dip characteristic of a single-photon source. The full dataset including the fits appears in the Supplementary Fig. 6.

Due to fit imperfections, at $t = 0$, there is still a narrow remnant of the antibunching dip. In the dip vicinity, however, we observe signal undulations originating from spin dynamics. We analyze them by performing a Fourier transform on the signal, presented in Fig. 7c. Aside from the low-frequencies concentration around 0, the Fourier spectrum reveals a clear component around $\pm 1.5$ GHz, marked in the figure by a green background. We then inverse-transform the Fourier spectrum only with these frequencies, reconstructing a filtered $g^{(2)}(\tau)$ function that exhibits clear oscillations (Fig. 7e). Fitting the data with the function $e^{-|t|/T_2^*}\cos(\omega|t| + \varphi)$, we extract the oscillation frequency $\omega$ and its coherence time $T_2^*$. Here, in contrast to the coherence time $T_2^*$ in the excited state as measured above, the measured coherence time $T_2^*$ is pump power dependent, since it is extracted from a constantly driven system. The real coherence time, as observed in an undisturbed system is expected to be higher, thus, all coherence times measured are lower bounds. To account for the errors, we fit the frequency concentrations in the Fourier spectrum with Gaussian peaks. These fits determine the exact frequency window we choose to inverse-transform, and their errors are then propagated to $\omega$ and $T_2^*$.

We repeat this procedure for different magnetic fields between 30 mT and 90 mT. Fig. 7 b,d,f display a similar measurement and analysis for $B_x = 90$ mT. Interestingly, the Fourier analysis in this case shows that a second prominent frequency component emerges (orange background in Fig. 7d). The two components, converted to an energy splitting are summarized in Fig. 7g as a function of the magnetic field. A linear fit to each dataset allows us to extract the $g$-factor of the oscillating spins. For the low frequency, we find $|g| = 0.568 \pm 0.004$ (orange datapoints). This value sufficiently agrees with the value $|g| = 0.593 \pm 0.002$ determined earlier from the lifetime oscillations (Fig. 5c). Therefore, we attribute it to the excited-level spin. By the method of elimination, we conclude that the other frequency component is associated with the ground-level spin. In this case, the linear fit in Fig. 7g determines $|g| = 2.876 \pm 0.044$ (green datapoints).

Another interesting difference between the excited- and ground-level spin precessions lies in their measured coherence times. In Fig. 7h, we plot them as a function of the applied magnetic field. A clear trend is observed: the coherence time of the excited-level spin is longer than that of the ground level-spin.

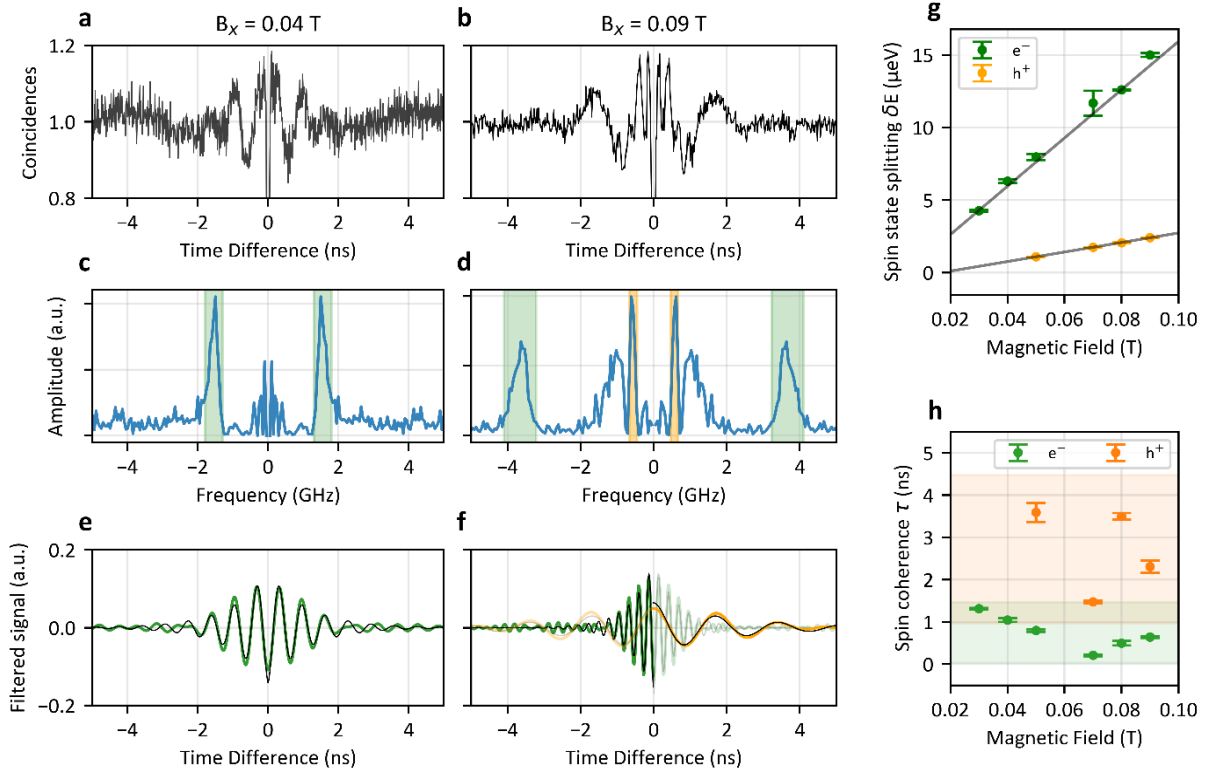

**Fig. 7. Polarization-resolved $g^{(2)}(\tau)$ of the $X^-$ under CW excitation**. **a,b** Two-photon coincidence histogram in the presence of in-plane magnetic fields with intensities 0.04 and 0.09 T, respectively. The data is normalized on a fit to the antibunching dip. **c,d** Fourier transform of the signal showing its spectrum in the frequency domain. A dominant high frequency in the spectrum is identified as the ground-level electron Larmor precession and marked by a green background. A low-frequency component, emerging at 0.09 T, is attributed to the excited-level Larmor precession which is governed by the hole (orange background). **e,f** Inverse-transform of the isolated frequency regions of interest. Fitting a decaying exponential envelope, we extract the dephasing time of the oscillations, $T_2^*$ (see text). **g** Electron and hole frequencies plotted against the applied magnetic field; linear fits reveal the corresponding absolute-value $g$-factors, $2.876 \pm 0.044$ and $0.568 \pm 0.004$. The error bars are propagated from gaussian fits to the selected frequency regions in (c) and (d) (see text). **h** Comparison of the oscillations decay times, interpreted as spin dephasing. Overall, the hole dephasing time is measured larger than the electron dephasing time, which is in line with the widespread premise.

At this point, we can strengthen our identification of the studied trion as the negative one, $X^-$. It is based on the following two independent arguments:

1. <u>$g$-factor comparison</u>. We associate the larger $g$-factor, 2.876, to the electron. We do that following a fairly established theoretical argument: an electron in III-IV semiconductors has an s-shell unit-cell symmetry, and therefore, its $g$-factor is expected to be isotropic. On the other hand, the hole has a p-shell symmetry, making its $g$-factor anisotropic. Specifically, its in-plane $g$-factor is expected to be smaller than the out-of-plane counterpart. Many studies have shown this phenomenon experimentally in InAs QDs emitting both at the 900 [36], and lately, at the 1500 nm range [33].
2. <u>Coherence time comparison</u>. The difference in the unit-cell symmetry between the two charge carriers affects also their interaction with the nuclei comprising the QD. The holes, having a p-symmetric unit-cell function, overlap less with these nuclei compared to the s-symmetric electrons. Consequently, a hole in a QD experiences a better effective isolation from the nuclei spin environment and has a longer coherence time. Studies on 900-nm-emitting QDs have reported a bound of 2.5 ns on the coherence time of the electron, and roughly 30 ns on the hole [29,37]. In this work, we measured the longer coherence time for the excited-level spin. The longest value, mentioned earlier in Section 2.3, was measured 6.5 ns. Based on this comparison, we argue that the excited-level spin is governed by the hole.

Since both reasonings agree, we conclude that the ground-level spin is the electron, and the excited-level spin is the $X^-$, governed by the hole.

## 3 Summary

To conclude, we have studied the time-dependent selection rules of the $X^0$ and $X^-$ optical transitions in QD emitting in the telecom C-band. We have shown that we can conceptually *write* the state of the $X^0$ using a picosecond laser pulse tuned to its quasi-resonance. We showed this by using the laser polarization to prepare the $X^0$ spin state in any arbitrary superposition. For the $X^-$, we measured its Larmor precession under externally applied in-plane magnetic field. We did that once in a lifetime measurement, and a second time in a polarization-sensitive $g^{(2)}(\tau)$ measurement. From these measurements, we extracted the electron and hole in-plane $g$-factors, $|g_e| = 2.876 \pm 0.044$ and $|g_h| = 0.593 \pm 0.002$. In combination with the coherence times measured for both charge carriers, including $T_2^* > 6.4$ ns for the hole, we were able to unambiguously determine the sign of the studied trion as negative. These tools are useful for a future quest in identifying a positive trion – a promising candidate to act as a photon entangler for scalable quantum information schemes.

## 4 Methods

*Sample*

We study InGaAs QDs grown by MOVPE using the Stranski-Krastanov method on a GaAs substrate. Typically, lattice mismatch between InAs and GaAs causes strain on the QDs and limits their size such that the confined excitonic energy lies within the near-IR wavelength range. Pushing this material system to the telecom range requires finding ways to reduce this strain. In our sample, a InGaAs metamorphic (MMB) buffer layer was used before the wetting layer growth, allowing the QDs to form larger. Details can be found in Ref [38]. The sample embeds the QDs in a 1 lambda planar cavity made from distributed Bragg reflectors, where the bottom mirror comprises 23 mirror pairs of AlAs/GaAs and the top mirror comprises 4 mirror pairs of $TiO_2/SiO_2$ which were sputter-deposited post growth. A sketch of the sample's layer structure can be found in the supplementary information.

*Excitation lasers*

Above bandgap: 1064 nm diode laser. CW tunable laser for PLE scan: Toptica[TD] CTL 1500. Pulsed laser for quasi-resonant excitations: APE picoEmerald[TD]. Pre-characterization (not shown) was performed using a ps-pulsed Ti:sapphire Tsunami[TD] laser from SpectraPhysics.

*Setup Polarization Control*

Polarization control was achieved using two pairs of liquid crystal variable retarders (LCVRs). One pair was used for the excitation path and the other for the detection. These retarders allowed us to rotate the Poincaré sphere of polarization from the lab frame to the QD frame. In the QD frame, the $H$ and $V$ polarizations are defined along the FSS splitting of the $X^0$. $D$ and $A$ are other orthogonal polarization pair in the sample plane, while the $R$ and $L$ are the circular polarization along the out-of-plane direction. For the measurements conducted on the $X^-$, polarization control in the detection was achieved using achromatic waveplates.

**Data availability.** Data underlying the results presented in this paper are not publicly available at this time but may be obtained from the authors upon reasonable request.

# References

1. Krantz, P., Kjaergaard, M., Yan, F., Orlando, T. P., Gustavsson, S. & Oliver, W. D. A quantum engineer's guide to superconducting qubits. *Appl. Phys. Rev.* **6,** 21318 (2019).

2. Monroe, C. *et al.* Programmable quantum simulations of spin systems with trapped ions. *Rev. Mod. Phys.* **93,** 25001 (2021).

3. Schäfer, F., Fukuhara, T., Sugawa, S., Takasu, Y. & Takahashi, Y. Tools for quantum simulation with ultracold atoms in optical lattices. *Nature Reviews Physics* **2,** 411–425 (2020).

4. Slussarenko, S. & Pryde, G. J. Photonic quantum information processing: A concise review. *Appl. Phys. Rev.* **6,** 41303 (2019).

5. Alexander, K. *et al.* A manufacturable platform for photonic quantum computing. Preprint at http://arxiv.org/abs/2404.17570 (2024).

6. Bartolucci, S. *et al.* Fusion-based quantum computation. *Nature Communications* **14,** 912 (2023).

7. Couteau, C. *et al.* Applications of single photons to quantum communication and computing. *Nature Reviews Physics* **5,** 326–338 (2023).

8. Akopian, N. *et al.* Entangled Photon Pairs from Semiconductor Quantum Dots. *Phys. Rev. Lett.* **96,** 130501 (2006).

9. Jayakumar, H., Predojević, A., Kauten, T., Huber, T., Solomon, G. S. & Weihs, G. Time-bin entangled photons from a quantum dot. *Nature Communications* **5,** 4251 (2014).

10. Lindner, N. H. & Rudolph, T. Proposal for Pulsed On-Demand Sources of Photonic Cluster State Strings. *Phys. Rev. Lett.* **103,** 113602 (2009).

11. Buterakos, D., Barnes, E. & Economou, S. E. Deterministic Generation of All-Photonic Quantum Repeaters from Solid-State Emitters. *Phys. Rev. X* **7,** 41023 (2017).

12. Borregaard, J., Pichler, H., Schröder, T., Lukin, M. D., Lodahl, P. & Sørensen, A. S. One-Way Quantum Repeater Based on Near-Deterministic Photon-Emitter Interfaces. *Phys. Rev. X* **10,** 21071 (2020).

13. Thomas, P., Ruscio, L., Morin, O. & Rempe, G. Fusion of deterministically generated photonic graph states. *Nature* **629,** 567–572 (2024).

14. Schwartz, I. *et al.* Deterministic generation of a cluster state of entangled photons. *Science* **354,** 434–437 (2016).

15. Cogan, D., Su, Z.-E., Kenneth, O. & Gershoni, D. Deterministic generation of indistinguishable photons in a cluster state. *Nature Photonics* **17,** 324–329 (2023).

16. Li, J.-P. *et al.* Multiphoton Graph States from a Solid-State Single-Photon Source. *ACS Photonics* **7,** 1603–1610 (2020).

17. Istrati, D. *et al.* Sequential generation of linear cluster states from a single photon emitter. *Nature Communications* **11,** 5501 (2020).

18. Coste, N. *et al.* High-rate entanglement between a semiconductor spin and indistinguishable photons. *Nature Photonics* **17,** 582–587 (2023).

19. Wang, J. *et al.* Multidimensional quantum entanglement with large-scale integrated optics. *Science* **360,** 285–291 (2018).

20. Thomas, P., Ruscio, L., Morin, O. & Rempe, G. Efficient generation of entangled multiphoton graph states from a single atom. *Nature* **608,** 677–681 (2022).

21. Joos, R. *et al.* Coherently and Incoherently Pumped Telecom C-Band Single-Photon Source with High Brightness and Indistinguishability. *Nano Letters* **24,** 8626–8633 (2024).

22. Kim, J. *et al.* Two-Photon Interference from an InAs Quantum Dot emitting in the Telecom C-Band. Preprint at https://arxiv.org/abs/2501.15970 (2025).

23. Dusanowski, Ł. *et al.* Optical charge injection and coherent control of a quantum-dot spin-qubit emitting at telecom wavelengths. *Nature Communications* **13,** 748 (2022).

24. Laccotripes, P., Müller, T., Stevenson, R. M., Skiba-Szymanska, J., Ritchie, D. A. & Shields, A. J. Spin-photon entanglement with direct photon emission in the telecom C-band. *Nature Communications* **15,** 9740 (2024).

25. Olbrich, F. *et al.* Polarization-entangled photons from an InGaAs-based quantum dot emitting in the telecom C-band. *Appl. Phys. Lett.* **111,** 133106 (2017).

26. Zeuner, K. D. *et al.* On-Demand Generation of Entangled Photon Pairs in the Telecom C-Band with InAs Quantum Dots. *ACS Photonics* **8,** 2337–2344 (2021).

27. Benny, Y. *et al.* Coherent Optical Writing and Reading of the Exciton Spin State in Single Quantum Dots. *Phys. Rev. Lett.* **106,** 40504 (2011).

28. Schwartz, I. *et al.* Deterministic Writing and Control of the Dark Exciton Spin Using Single Short Optical Pulses. *Phys. Rev. X* **5,** 11009 (2015).

29. Cogan, D. *et al.* Depolarization of Electronic Spin Qubits Confined in Semiconductor Quantum Dots. *Phys. Rev. X* **8,** 41050 (2018).

30. Benny, Y., Kodriano, Y., Poem, E., Khatsevitch, S., Gershoni, D. & Petroff, P. M. Two-photon photoluminescence excitation spectroscopy of single quantum dots. *Phys. Rev. B* **84,** 75473 (2011).

31. Winik, R. *et al.* On-demand source of maximally entangled photon pairs using the biexciton-exciton radiative cascade. *Phys. Rev. B* **95,** 235435 (2017).

32. Poem, E. & Gershoni, D. Radiative Cascades in Semiconductor Quantum Dots. In *Handbook of Luminescent Semiconductor Materials* (eds Bergman, L. & McHale, J. L.), 321–364 (CRC Press, New York, 2020).

33. Belykh, V. V. *et al.* Electron and hole g factors in InAs/InAlGaAs self-assembled quantum dots emitting at telecom wavelengths. *Phys. Rev. B* **92,** 165307 (2015).

34. Belykh, V. V. *et al.* Large anisotropy of electron and hole g factors in infrared-emitting InAs/InAlGaAs self-assembled quantum dots. *Phys. Rev. B* **93,** 125302 (2016).

35. Coste, N. *et al.* Probing the dynamics and coherence of a semiconductor hole spin via acoustic phonon-assisted excitation. *Quantum Science and Technology* **8,** 25021 (2023).

36. Witek, B. J., Heeres, R. W., Perinetti, U., Bakkers, E. P. A. M., Kouwenhoven, L. P. & Zwiller, V. Measurement of the g-factor tensor in a quantum dot and disentanglement of exciton spins. *Phys. Rev. B* **84,** 195305 (2011).

37. Bechtold, A. *et al.* Three-stage decoherence dynamics of an electron spin qubit in an optically active quantum dot. *Nature Physics* **11,** 1005–1008 (2015).

38. Sittig, R. *et al.* Thin-film InGaAs metamorphic buffer for telecom C-band InAs quantum dots and optical resonators on GaAs platform. *Nanophotonics* **11,** 1109–1116 (2022).

**Author contributions.** M.J. and P.M. provided the sample. G.P., J.M., R.J, and M.H. built the spectroscopy setup, conducted the measurements, and carried out the data analysis. G.P. and J.M. wrote the paper with input from all authors. S.L.P, P.M, S.H., and T.H-L guided the work and acquired funding.

**Funding.** The authors are grateful for financial support within the projects QR.X (FKZ: 16KISQ010 and FKZ: 16KISQ013), QR.N (FKZ: 16KIS2207), PhotonQ (FKZ: 13N15759), and QECS (FKZ: 13N16272) funded by the German Ministry for Research and Education (BMBF). We also acknowledge financial support by the State of Bavaria and the German research foundation (DFG) under the project Nr. INST 93/1007-1 LAGG.

**Competing interests.** The authors declare no competing interests.